\documentclass[preprint,preprintnumbers,amsmath,amssymb]{revtex4}
\usepackage{graphicx}
\usepackage{epstopdf}
\usepackage{dcolumn}
\usepackage{bm}
\usepackage{color}

\begin{document}


\title{The graphene/$n$-Ge(110) interface: structure, doping, and electronic properties}

\author{Julia Tesch,$^{1}$ Fabian Paschke,$^{1}$ Mikhail Fonin,$^{1}$ Marko Wietstruk,$^{2}$ Stefan B\"ottcher,$^{2}$ Roland J. Koch,$^{3}$ Aaron Bostwick,$^{3}$ Chris Jozwiak,$^{3}$ Eli Rotenberg,$^{3}$ Anna Makarova,$^{4}$ Beate Paulus,$^{5}$ Elena Voloshina,$^{6,}$\footnote{Corresponding author. E-mail: voloshina@shu.edu.cn} and Yuriy Dedkov$^{6,}$\footnote{Corresponding author. E-mail: dedkov@shu.edu.cn}}

\affiliation{$^1$Fachbereich Physik, Universit\"at Konstanz, 78457 Konstanz, Germany}
\affiliation{$^2$SPECS Surface Nano Analysis GmbH, Voltastra\ss e 5, 13355 Berlin, Germany}
\affiliation{$^3$Advanced Light Source, E. O. Lawrence Berkeley National Laboratory, Berkeley, CA, 94720, USA}
\affiliation{$^4$Institut f\"ur Festk\"orperphysik, Technische Universit\"at Dresden, 01062 Dresden, Germany}
\affiliation{$^5$Institut f\"ur Chemie und Biochemie, Freie Universit\"at Berlin, Takustra\ss e 3, 14195 Berlin, Germany}
\affiliation{$^6$Department of Physics and International Centre for Quantum and Molecular Structures, Shanghai University, Shangda Road 99, 200444 Shanghai, China}

\date{\today}

\begin{abstract}
The implementation of graphene in semiconducting technology requires the precise knowledge about the graphene-semiconductor interface. In our work the structure and electronic properties of the graphene/$n$-Ge(110) interface are investigated on the local (nm) and macro (from $\mu\mathrm{m}$ to mm) scales via a combination of different microscopic and spectroscopic surface science techniques accompanied by density functional theory calculations. The electronic structure of freestanding graphene remains almost completely intact in this system, with only a moderate $n$-doping indicating weak interaction between graphene and the Ge substrate. With regard to the optimization of graphene growth it is found that the substrate temperature is a crucial factor, which determines the graphene layer alignment on the Ge(110) substrate during its growth from the atomic carbon source. Moreover, our results demonstrate that the preparation routine for graphene on the doped semiconducting material ($n$-Ge) leads to the effective segregation of dopants at the interface between graphene and Ge(110). Furthermore, it is shown that these dopant atoms might form regular structures at the graphene/Ge interface and induce the doping of graphene. Our findings help to understand the interface properties of the graphene-semiconductor interfaces and the effect of dopants on the electronic structure of graphene in such systems. 

\end{abstract}

\maketitle

\section{Introduction}

Systems on the basis of graphene (gr) layers on different supports - metallic, semiconducting, or insulating - are the focus of intensive fundamental and technology-related studies~\cite{Tetlow:2014kd,Dedkov:2015kp,Wang:2016ca,Rezapour:2017gk}. Many of them aim at the integration of the obtained graphene layers in next generation micro(nano)-electronic and spintronic systems. However, different graphene-substrate systems have various drawbacks, which can limit their adaptation in technological processes. For example, graphene synthesis on polycrystalline Cu foil is used for the fabrication of the dozens-inches-sized graphene, which can then be transferred onto the desired polymer or insulator support~\cite{Bae:2010} and subsequently be used for the preparation of touch screens, sensors, etc.~\cite{Ryu:2014fo,Park:2015hn,Sahatiya:2016er} Nevertheless, as was recently found, the level of residual metallic contamination does not allow for the adaptation of this graphene synthesis procedure on metallic support to modern semiconductor technology~\cite{Ambrosi:2014gl,Lupina:2015je}. Growth on insulating oxide surfaces requires relatively high temperatures of about or more than $1000^\circ$\,C and the quality of the synthesized graphene layer remains to be improved~\cite{Gaddam:2011kz,Sun:2014fia}.

Taking into account the previously described drawbacks, the direct growth of graphene on semiconducting surfaces is the most promising way for graphene implementation in modern semiconductor technology. The natural choice of SiC as a substrate for the graphene growth is limited by the relatively high price of the respective substrate (i.\,e. large scale wafers) as well as the high temperature used during synthesis~\cite{Emtsev:2009,Kruskopf:2016eu}. Alternatively to the previously described approaches of graphene integration in semiconductor technology, the new method of graphene growth on semiconducting Ge surfaces was recently developed~\cite{Lee:2014dv}. This approach might be technologically relevant as Ge(001)/Si(001) and Ge(110)/Si(110) epilayers are readily available and used in present day semiconducting applications. 

Further experimental and theoretical studies on the gr/Ge system demonstrate that although the Ge(001) surface is the most suitable for the processing, it undergoes facetting upon graphene synthesis and the formation of the Ge(107) facets was determined by means of different microscopy techniques~\cite{Jacobberger:2015de,Kiraly:2015kaa,Pasternak:2016ec,Lukosius:2016ce,Dabrowski:2017gr}. For gr/Ge(110) and gr/Ge(111) no such rearrangement of the underlying Ge surface was observed~\cite{Lee:2014dv,Kiraly:2015kaa}, which makes these gr/Ge interfaces more suitable for the further applications. The electronic structure of the gr/Ge interfaces was partially addressed in recent studies, where contradictory results were obtained for graphene grown on Ge(001)/Si(001) epilayers by means of molecular-beam epitaxy (MBE) from the atomic-carbon source~\cite{Dabrowski:2016im} and chemical-vapour deposition (CVD) using the CH$_4$/Ar/H$_2$ mixture~\cite{Dabrowski:2017gr}. Graphene was found to be either $p$-doped with a position of the Dirac point at $E_D-E_F=+0.185$\,eV~\cite{Dabrowski:2016im} or slightly $n$-doped with $E_D-E_F=-0.05...-0.1$\,eV~\cite{Dabrowski:2017gr}, respectively (here $E_F$ is the Fermi energy and $E_D$ is the Dirac point energy). The effect of possible dopants has not been investigated up to now.

Here we present our studies of the structure and electronic properties of graphene prepared by means of MBE from an atomic-carbon source on $n$-doped (Sb) Ge(110). These studies were performed via a combination of different microscopic and spectroscopic methods accompanied by density functional theory (DFT) calculations. We found that the substrate temperature used during graphene growth is the crucial factor which determines the alignment of the graphene layer on Ge(110). Our results demonstrate that graphene on $n$-Ge(110) is almost free-standing and slightly $n$-doped due to the segregation of Sb-dopants at the interface during the preparation procedure. These findings are supported by the present DFT results allowing to identify the source of graphene doping at the gr/Ge interfaces.

\section{Results and discussion}

\begin{figure}[t]
\includegraphics[width=0.9\textwidth]{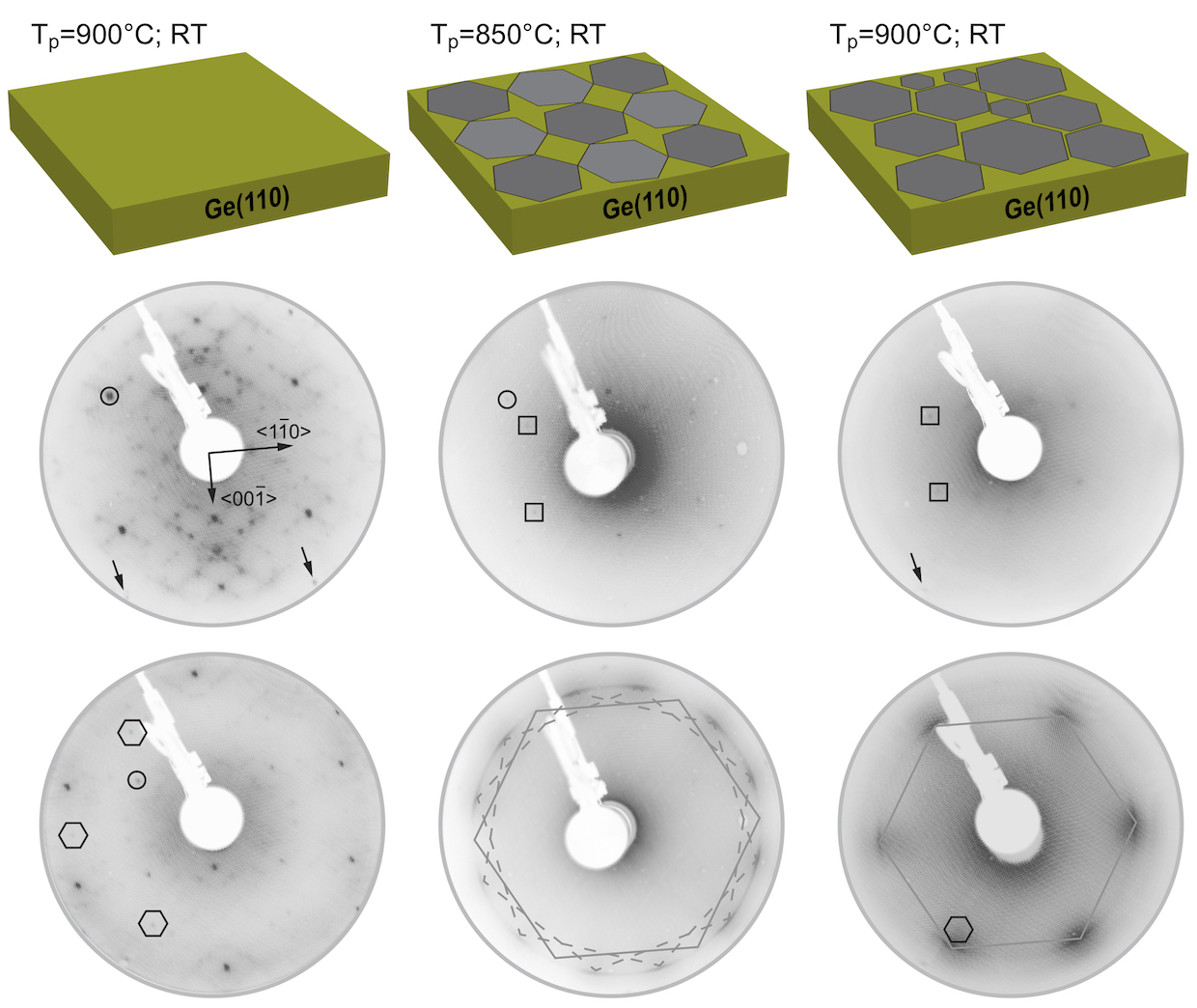}\\
\caption{LEED images of Ge(110) (left column) and gr/Ge(110) prepared at two different substrate temperatures of $T=850^\circ$\,C (Sample A, middle column) and $T=900^\circ$\,C (Sample B, right columns). Images were acquired at the primary electron energy of $40$\,eV (middle row) and $75$\,eV (bottom row). Small circles, rectangles, and hexagons mark the same LEED spots of Ge(110) before and after graphene growth and are used to build the crystallographic model of the gr/Ge(110) interface. Dashed- and solid-line big hexagons mark LEED spots for the two-domain (Sample A) and single-domain (Sample B) graphene, respectively.}
\label{grGe_growthLEED}                                                                                            
\end{figure}

The quality and the crystallographic structure of the grown graphene layers on $n$-Ge(110) were investigated by means of low-energy electron diffraction (LEED) and scanning tunneling microscopy (STM). Fig.~\ref{grGe_growthLEED} shows LEED images of the clean Ge(110) surface (left column) as well as of the gr/Ge(110) interface prepared at two different substrate temperatures used during carbon deposition: $850^\circ$\,C (Sample A, middle column) and $900^\circ$\,C (Sample B, right column). The LEED image of Ge(110) demonstrates diffraction spots, which can be assigned to the $c(8\times10)$-reconstructed steps of Ge(110) separated by $\{17\,15\,1\}$ facets~\cite{Olshanetsky:1977da,Ichikawa:2004cw,Mullet:2014bd}. Deposition of carbon on Ge(110) kept at $850^\circ$\,C followed by the sample cooling (Sample A) leads to the formation of two sets of diffraction spots of hexagonal symmetry (Fig.~\ref{grGe_growthLEED}, middle column). As was shown earlier in Ref.~\citenum{Tesch:2017gm}, these spots can be assigned to two types of graphene domains, which are misoriented by $\pm15^\circ$ with respect to the graphene single-domain orientation observed earlier~\cite{Lee:2014dv,Kiraly:2015kaa,Rogge:2015hl,Dai:2016jm} and in the present work (see below).

The increase of the Ge substrate temperature during carbon deposition up to $900^\circ$\,C leads to single-domain graphene growth on Ge(110) (Fig.~\ref{grGe_growthLEED}, right column). The analysis of the LEED images of Ge(110) and gr/Ge(110) allows to conclude that for the single-domain sample (Sample B) the armchair direction of the graphene lattice is parallel to the $<1\bar{1}0>$ direction of Ge(110), similar to what was found in previous studies~\cite{Rogge:2015hl,Dai:2016jm}. The formation of a graphene layer on Ge(110) leads to a lifting of the original $c(8\times10)$ reconstruction of Ge(110), but this surface undergoes further restructuring (middle row in Fig.~\ref{grGe_growthLEED}). Our preliminary analysis shows that the Ge surface underneath graphene rearranges into a $(n\times2)$ structure, however, further studies in this direction are necessary.

\begin{figure}[t]
\includegraphics[width=0.9\textwidth]{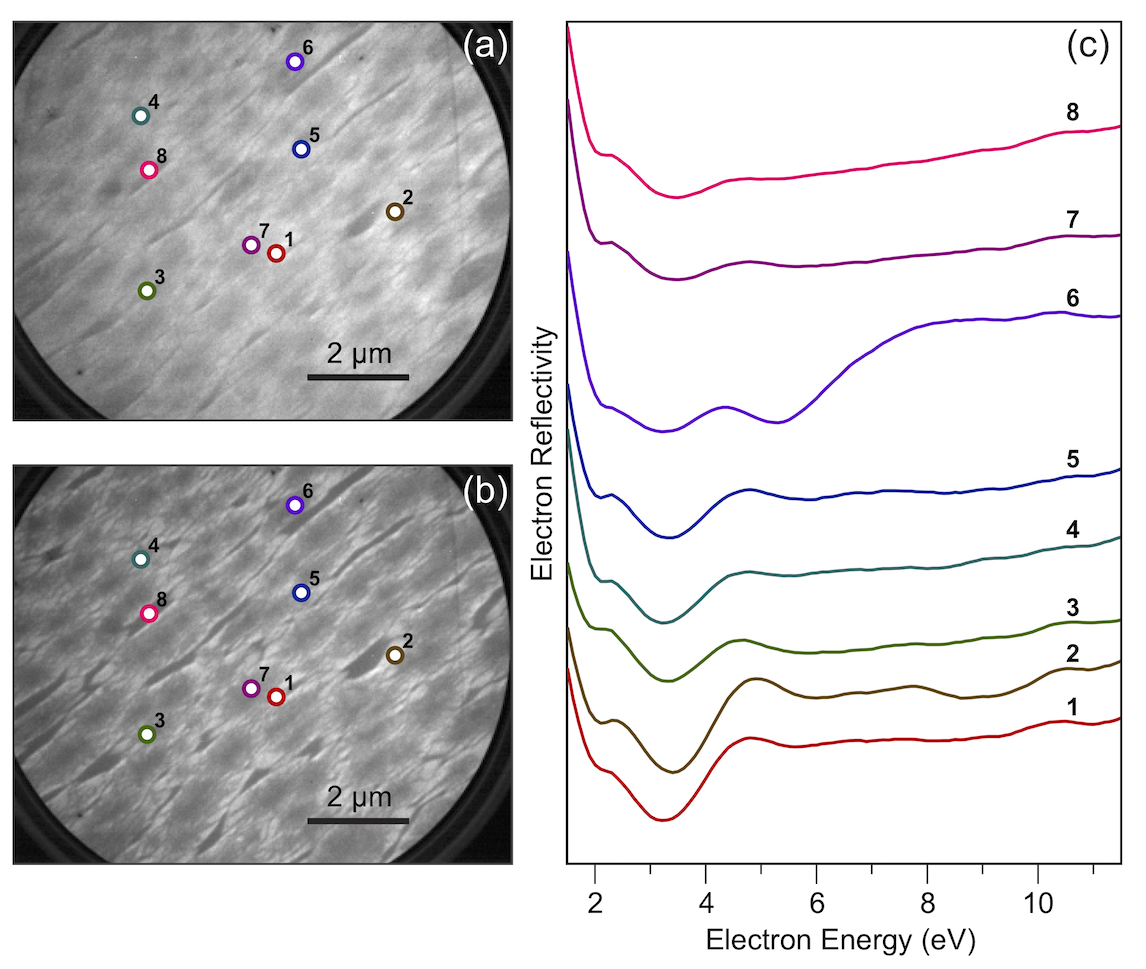}\\
\caption{(a,b) LEEM images of gr/Ge(110) (Sample A) recorded at the electron energy of $E_{vac}+2.82$\,eV and $E_{vac}+4.80$\,eV, respectively. (c) Electron reflectivity as a function of electron energy extracted for several places of gr/Ge(110) and marked by the respective digits in (a,b).}
\label{grGe_mLEEM}                                                                                            
\end{figure}

The morphology and the thickness distribution of the formed graphene layer on Ge(110) were studied by means of low-energy electron microscopy (LEEM) and the results for Sample A are presented in Fig.~\ref{grGe_mLEEM}. This figure shows two LEEM images recorded at (a) $E_{vac}+2.82$\,eV and (b) $E_{vac}+4.80$\,eV with a field of view (FOV) of $9.7\,\mu\mathrm{m}$. The set of images collected in the energy range of $-1\,...\,20$\,eV with respect to $E_{vac}$ is used for the extraction of the space resolved electron reflectivity spectra and a series of such spectra is presented in (c) for the different positions marked in the LEEM images. According to Refs.~\citenum{Hibino:2008ki,Ohta:2008} the number of low-energy minima for the electron reflectivity curve corresponds to the number of graphene layers on the substrate as the conduction band of graphite along the $\Gamma-\mathrm{A}$ direction of the Brillouine zone is quantized due to the finite thickness of the film. From the spectra presented in panel (c) we conclude that prepared graphene layer on Ge(110) is uniform on the $\mu\mathrm{m}$-scale with a thickness of $1$\,ML (except for very small areas). 

\begin{figure}[t]
\includegraphics[width=1\textwidth]{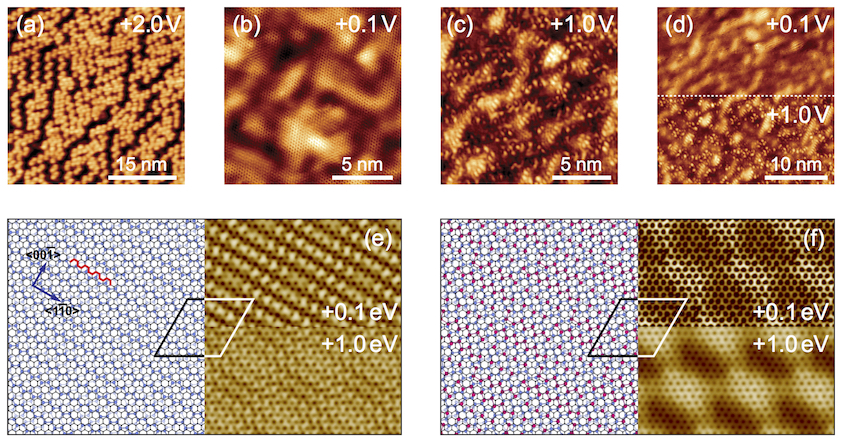}\\
\caption{STM images of (a) a clean Ge(110) surface and (b-d) gr/Ge(110) (Sample B) at different bias voltages (marked in every image). Tunneling parameters: (a) $I_T=1$\,nA, (b-d) $I_T=400$\,pA. Bottom row presents structural models and the respective simulated STM images for (e) gr/Ge(110) and (f) gr/Sb/Ge(110) interfaces. Black, blue, and red spheres are C, Ge, and Sb atoms, respectively.}
\label{grGe_STM_bias}                                                                                            
\end{figure}

The ordering and structure of the gr/$n$-Ge(110) interface on the nm-scale was investigated by means of STM at different bias voltages. Such experiments allow to discriminate between graphene and Ge(110) contributions in the imaging and to identify the relative alignment of atoms in the graphene layer and the Ge(110) substrate. Fig.~\ref{grGe_STM_bias} shows STM images of (a) clean Ge(110) and (b-d) gr/Ge(110) (Sample B) collected at different bias voltages marked in every image. In image (d) the bias voltage was changed \textit{on-the-fly} from $+1.0$\,V to $+0.1$\,V in order to trace the atomic alignment as well as the locations of the impurities in the sample. Similar to Ref.~\citenum{Dai:2016jm} and to the present LEED measurements (see Fig.~\ref{grGe_growthLEED}), we found that the armchair direction of the graphene lattice is parallel to the $<1\bar{1}0>$ direction of Ge(110) as shown in the left-hand side of Fig.~\ref{grGe_STM_bias}(e), where the DFT optimized crystallographic structure of the gr/Ge(110) interface is presented (detailed discussion of the DFT results is presented below). The respective simulated STM images of such gr-Ge interface calculated for the experimentally used bias voltages are shown on the right-hand side of the same panel. One can clearly see that experimentally obtained STM images of gr/Ge(110) demonstrate pronounced areas of high electron density, which are imaged as bright patches. Variation of the bias voltage between $\pm0.1$\,V and $\pm1.0$\,V allows to assign them to agglomerations of the impurity atoms which segregate at the gr/Ge(110) interface and contribute strongly to the imaging contrast. For the present samples, these impurities are Sb atoms, which can segregate during the sample preparation routine (sample sputtering and high-temperature annealing) and, as was shown in our previous~\cite{Tesch:2017gm} and present experiments, they can define the doping level of the formed graphene layer on Ge(110). The DFT optimized crystallographic structure of the gr/Sb/Ge(110) interface with an ordered Sb layer and the respective calculated STM images are shown in Fig.~\ref{grGe_STM_bias}(f). Experimental and theoretically calculated STM images are different with respect to the positions of the Sb atoms, disordered agglomerations for the experiment and ordered for the model used in our DFT calculations. Nevertheless we observed a reasonable agreement between the two images. Moreover, the inclusion of the correct concentration of Sb atoms at the gr-Ge interface in our model ($27$ Sb atoms per $(9 \times 9)$ graphene supercell) leads to the valid description of the doping level of graphene obtained from scanning tunneling spectroscopy (STS) and angle-resolved photoelectron spectroscopy (ARPES) measurements as well as the graphene corrugation in the STM experiment (detailed discussion of the DFT results is presented below). 

\begin{figure}[t]
\includegraphics[width=1\textwidth]{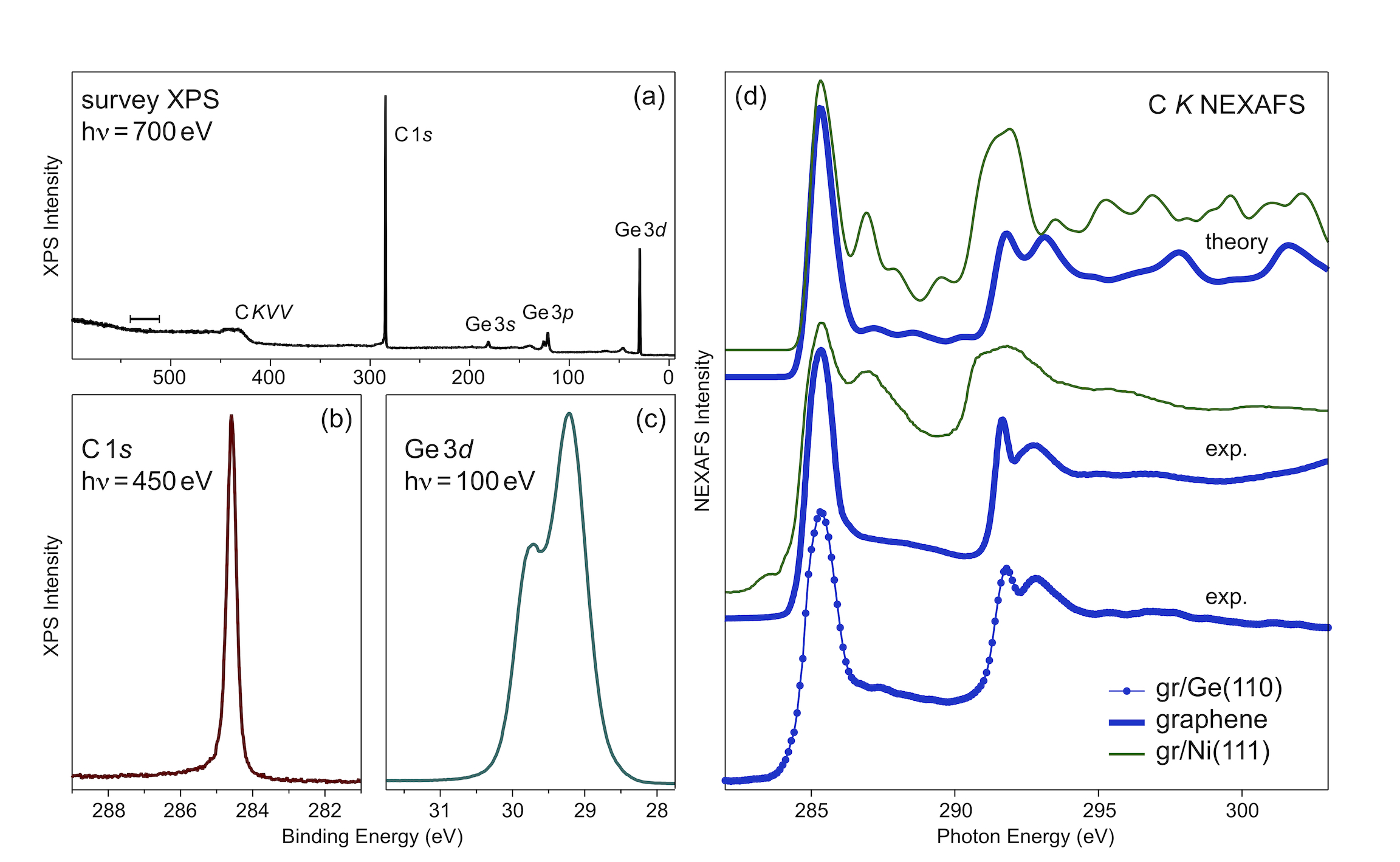}\\
\caption{(a-c) XPS spectra of gr/$n$-Ge(110): survey, C\,$1s$, and Ge\,$3d$. Photon energies used in every measurement are marked in every panel. A vertical bar in (a) marks the energy region where O\,$1s$ can be expected in case of a contaminated sample. (d) Experimental C $K$-edge NEXAFS spectra of gr/$n$-Ge(110), graphene, and gr/Ni(111). The respective theoretical spectra for the last two samples are shown at the top of the panel.}
\label{grGe_XPS_NEXAFS}
\end{figure}

\begin{figure}[t]
\includegraphics[width=1\textwidth]{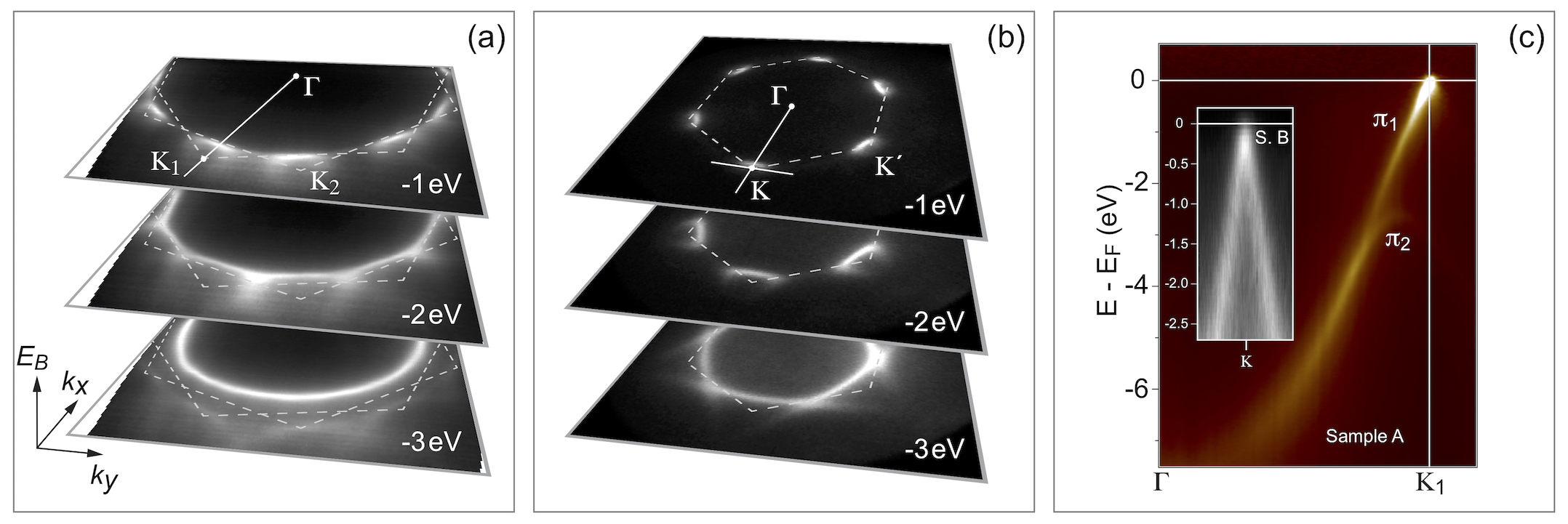}\\
\caption{Constant energy cuts of the ARPES intensity for (a) two-domain (Sample A) and (b) single-domain (Sample B) gr/$n$-Ge(110). Dashed-line hexagons mark respective hexagonal Brillouin zones of graphene for both samples. CECs are presented for binding energies of $1$\,eV, $2$\,eV, and $3$\,eV and were extracted from the complete 3D data sets of the ARPES intensity. (c) ARPES intensity map presented along the $\Gamma-\mathrm{K}_1$ direction (marked in (a)) of the graphene-derived Brillouin zone of one of the graphene domains in Sample A. Inset of (c) shows the photoemission intensity cut along the direction perpendicular to $\Gamma-\mathrm{K}$ (marked in (b)) for sample B. All ARPES data were collected at $T=100$\,K with a photon energy of $h\nu=100$\,eV.}
\label{grGe_ARPES}
\end{figure}

Systematic photoelectron spectroscopy studies of the gr/$n$-Ge(110) system on the hundred-$\mu\mathrm{m}$-scale were performed in a series of beamtimes at the BESSY\,II synchrotron facility and these results are compiled in Figs.~\ref{grGe_XPS_NEXAFS} and \ref{grGe_ARPES}. The cleanliness of the sample after annealing at $800^\circ$\,C in ultra-high vacuum (UHV) conditions as well as the chemical state of the elements in the system were probed by means of x-ray photoelectron spectroscopy (XPS) (Figs.~\ref{grGe_XPS_NEXAFS}(a-c)), which demonstrate the absence of oxygen contamination at the interface: the possible energy range for the O\,$1s$ line is marked by the horizontal bar in panel (a) and the absence of any trace of GeO$_x$ in the surface sensitive measurements of the Ge\,$3d$ line confirms this result. A single sharp C\,$1s$ line measured for gr/Ge(110) indicates the homogeneous $sp^2$ structure of the formed graphene layer. A comparison of the binding energy of this line at $(284.56\pm0.02)$\,eV with the respective positions for other epitaxial graphene-based systems~\cite{Schroder:2016eb} indicates the $n$-doping of graphene. This result is opposite to the previous experiments, where $p$-doped graphene was observed: by means of ARPES, DFT and transport measurements for the Ge-intercalated gr/SiC system~\cite{Emtsev:2011fo,Kaloni:2012be,Baringhaus:2015bz} and by means of nano-ARPES for graphene layers grown via CVD and MBE methods on Ge(001)/Si(001) epilayers~\cite{Dabrowski:2016im}. However, recent ARPES experiments on similarly CVD-prepared gr/Ge(001)/Si(001) samples demonstrate the slight $n$-doping of graphene with a position of the Dirac point of $E_D-E_F=-0.05\,...\,-0.1$\,eV.

Figure~\ref{grGe_XPS_NEXAFS}(d) shows the results of the C $K$-edge near-edge x-ray absorption fine structure (NEXAFS) studies of the gr/$n$-Ge(110) system. In this method, the photon energy is scanned around the value corresponding to the binding energy of the C\,$1s$ core level, leading to the excitation of this electron to the unoccupied valence band states of graphene. This method is element specific, giving the information about the partial density of states above $E_F$ as well as allowing to determine the spatial orientation of these states via the so-called search-light-like effect. The analysis of the angle-dependence of the NEXAFS signal allows to assign intensity peaking at $285.3$\,eV to the excitation of the $1s$ electron onto the $\pi^*$ unoccupied states of graphene and intensity in the range of $291-295$\,eV to the $1s\rightarrow\sigma^*$ transition. These results for gr/Ge(110) can be compared to the experimental and theoretical NEXAFS spectra of graphene and gr/Ni(111)~\cite{Voloshina:2013cw}. The shape of both $1s\rightarrow\pi^*$ and $1s\rightarrow\sigma^*$ NEXAFS lines as well as the distance between them can be taken as a measure of the interaction strength between graphene and substrate. In the case of strongly interacting graphene-metal systems, like gr/Ni(111) or gr/Rh(111)~\cite{Preobrajenski:2008,Weser:2010}, where orbital mixing of the graphene- and substrate-derived states is observed, the shape of both NEXAFS transition lines is strongly disturbed compared to the ones for free-standing graphene. The distance between the two main peaks is also reduced due to the partial $sp^2-\mathrm{to}-sp^3$ rehybridization observed in graphene in these systems. For free-standing graphene or weakly interacting graphene-substrate systems, like gr/Pt(111) or gr/Al(111)~\cite{Preobrajenski:2008,Voloshina:2011NJP}, the C $K$-edge NEXAFS spectra are very similar to the one obtained in our experiments for the gr/Ge(110) system, proving the weak interaction between graphene and Ge substrate without valence orbital mixing at the interface.

The electronic structure below $E_F$ of both gr/Ge(110) samples, A and B, was studied by means of ARPES (Fig.~\ref{grGe_ARPES}) (see also Movies S1 and S2 in the Supplementary Material). Panels (a) and (b) show stacks of the constant energy cut (CEC) ARPES intensity maps extracted from the complete 3D PES intensity data sets, $I(E_B,k_x,k_y)$, for the two-graphene-domains (Sample A) and for the single-graphene-domain (Sample B) layers on Ge(110), respectively. These maps are presented for $E-E_F=-1$\,eV, $-2$\,eV, and $-3$\,eV. As is well-known for free-standing graphene, in the vicinity of $E_D$ (which in this case coincides with $E_F$) the energy of the electronic states of graphene depends linearly on the wave-vector $k$. These energy bands form the so-called Dirac cones at the six $\mathrm{K}$-points in the corners of the hexagonal Brillouin zone of graphene. For Sample A, in agreement with the above presented LEED data (Fig.~\ref{grGe_growthLEED}, middle column), we observed $12$ equivalent graphene ARPES emission spots centred at the $12$ $\mathrm{K}$-points belonging to two Brillouin zones and originating from two equivalent, but misoriented graphene domains on Ge(110) present for this type of sample. Two equivalent points in the hexagonal Brillouin zones for these two domains are marked by $\mathrm{K}_1$ and $\mathrm{K}_2$, respectively. In case of Sample B, where LEED shows solely one set of hexagonal spots for the single-graphene-domain layer on Ge(110) (Fig.~\ref{grGe_growthLEED}, right-hand column), only one set of the ARPES emission spots centred at the $6$ $\mathrm{K}$-points of the hexagonal Brillouin zone is observed, confirming the high-quality of the studied samples.

Figure~\ref{grGe_ARPES}(c) shows the ARPES intensity map along the $\Gamma-\mathrm{K}_1$ direction of the Brillouin zone (marked in the upper slice of (a)) corresponding to one of the graphene domains of Sample A. The dispersions of the graphene-derived $\pi$ bands for the respective graphene domains are clearly resolved: $\pi_1$ is going towards $E_F$ and $\pi_2$ is bent downwards at $\approx3$\,eV binding energy, as the dispersion for this band is related approximately to the  $\Gamma-\mathrm{M}$ direction of the graphene Brillouin zone for the second graphene domain of Sample A. The inset of panel (c) shows the ARPES intensity map taken as a cut at the $\mathrm{K}$-point through the data set for Sample B along the direction perpendicular to $\Gamma-\mathrm{K}$ (marked in the upper CEC of (b)). Although the ARPES intensity is slightly blurred along these directions for both samples (see also the respective LEED images in Fig.~\ref{grGe_growthLEED}), our estimation for the Dirac point position gives $E_D-E_F=-210\pm50$\,meV, i.\,e. graphene is $n$-doped in the studied system. The linear fit of the ARPES dispersion in the vicinity of $E_F$ for Sample B yields Fermi velocity of $v_F=(1.38\pm0.15)\times10^6$\,m/s, which is in agreement with previous nano-ARPES experiments~\cite{Dabrowski:2016im}. These results are consistent with our previous preliminary data~\cite{Tesch:2017gm} as well as with present results (see also discussion below). 

\begin{figure}[t]
\includegraphics[width=1\textwidth]{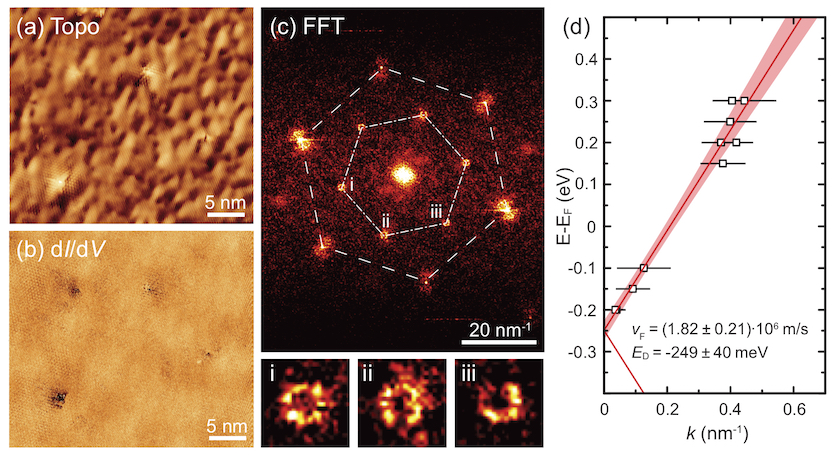}\\
\caption{(a) STM and (b) $dI/dV$ images of gr/Ge(110) acquired at $V_T=+200$\,mV and $I_T=800$\,pA. (c) FFT image of the $dI/dV$ map presented in (b). Large dashed-line hexagon marks the reciprocal lattice of graphene. Small point-dashed-line hexagon marks the Brillouin zone of graphene, where intervalley scattering spots are located at the ($\sqrt{3} \times \sqrt{3}$)$R30^\circ$-positions (marked with (i), (ii), and (iii) and their zoom are shown at the bottom of the panel). (d) Energy dispersion $E(k)$ of the electronic states of graphene on $n$-Ge(110) in the vicinity of the $\mathrm{K}$-point (open rectangles are experimental points and solid line is the linear fit with $E_D$ and $v_F$ parameters marked in the panel.}
\label{grGe_dIdV}
\end{figure}

The previous findings for the gr/$n$-Ge(110) system are confirmed by local nm-scale STM/STS experiments. A large scale atomically resolved STM and the respective $dI/dV$ images of gr/Ge(110) (Sample B) acquired at $V_T=+200$\,mV are shown in Fig.~\ref{grGe_dIdV}(a) and \ref{grGe_dIdV}(b), respectively. A fast Fourier transform (FFT) of the $dI/dV$ image is shown in Fig.~\ref{grGe_dIdV}(c), where several characteristic features can be identified. Firstly, the spots, which are in the corners of the big dashed-line hexagon, are assigned to the reciprocal lattice of graphene and they originate from the atomic resolution imaging in STM and during $dI/dV$ mapping.

The most interesting features are the ring-like structures located at the positions of the $(\sqrt{3}\times\sqrt{3})R30^\circ$ points with respect to graphene's atomic reciprocal structure (marked as a dot-dashed-line hexagon in Fig.~\ref{grGe_dIdV}(c)). Spots are marked as (i), (ii), (iii) and their zooms are presented in the bottom panels of (c). These features are assigned to the intervalley scattering between graphene-related valence band states around the $\mathrm{K}$ and $\mathrm{K'}$ points~\cite{Rutter:2007ep,Simon:2011dv,Mallet:2012ib,Leicht:2014jy}. The radius of these rings is $2k$, where $k$ is the wave vector of the Dirac particles at an energy $E$ relative to $E_F$ and it is measured with respect to the $\mathrm{K}$-point of the graphene Brillouin zone. An analysis of a series of such $dI/dV$ maps measured at different bias voltages $V_T$ allows one to identify particular scattering vectors between different electronic states in the Brillouin zone at a fixed energy, $E = eV_T$, and to plot the energy dispersion of the charge carriers $E(k)$ around $E_F$. The results of these studies for gr/$n$-Ge(110) are presented in Fig.~\ref{grGe_dIdV}(d) where experimental data points as well as the respective linear fit are shown. According to these data graphene in the gr/$n$-Ge(110) system is $n$-doped with the position of the Dirac point of $E_D-E_F=-249\pm40$\,meV and the extracted Fermi velocity is $v_F=(1.82\pm0.21)\times10^6$\,m/s.

The relatively high value of the Fermi velocity measured at $100$\,K (ARPES) and $10$\,K (STM/STS) for graphene on Ge(110) can be assigned to the presence of the semiconducting substrate, which leads to a reduced screening of the electron-electron interaction in graphene compared to the fully screened case. If we assume the ideal case of fully screened electron-electron interaction in graphene, then the dielectric constant $\varepsilon=\infty$ and within the local-density approximation (LDA) in DFT one obtains $v_F(\mathrm{LDA})=0.85\times10^6$\,m/s~\cite{Trevisanutto:2008fs}. The growth of graphene on a semiconducting substrate leads to effective dielectric constant of $\varepsilon=(\varepsilon_{\mathrm{s}}+\varepsilon_{\mathrm{v}})/2$ and to the Fermi velocity renormalization~\cite{Park:2009ku} ($\varepsilon_{\mathrm{s}}$ and $\varepsilon_{\mathrm{v}}$ are the dielectric constants of substrate and vacuum, respectively). Our result for gr/Ge(110) is in line with the recently reported high values of the Fermi velocity measured in gr/SiC, gr/$h$-BN, gr/quartz, and gr/SrTiO$_3$ systems~\cite{Hwang:2012he,Ryu:2017dx}. The increase of the Fermi velocity in graphene on Ge(110) at low temperatures ($10$\,K vs. $100$\,K) is assigned to the slight decrease of $\varepsilon_{\mathrm{Ge}}$ with temperature decrease~\cite{Smakula:1972jl,Samara:1983bb}.

The above presented experimental results were analyzed in the framework of the DFT approach. Several models of the gr/$n$-Ge(110) interface with inclusion of Sb dopants were considered: (i) clean interface, (ii) Sb atoms inside the Ge slab, (iii) different concentrations of Sb atoms at the gr/Ge interface. In all cases, the atomic positions were allowed to relax according to the procedure described in Methods until the equilibrium geometry was reached and then other characteristics, such as STM images or the band structure, which can be compared with the experimental data, were calculated. The resulting graphene corrugation, main distances, graphene adsorption energies, as well as the position of the Dirac point for all systems are summarized in Table S3 of the Supplementary Material. 

\begin{figure}[t]
\includegraphics[width=1\textwidth]{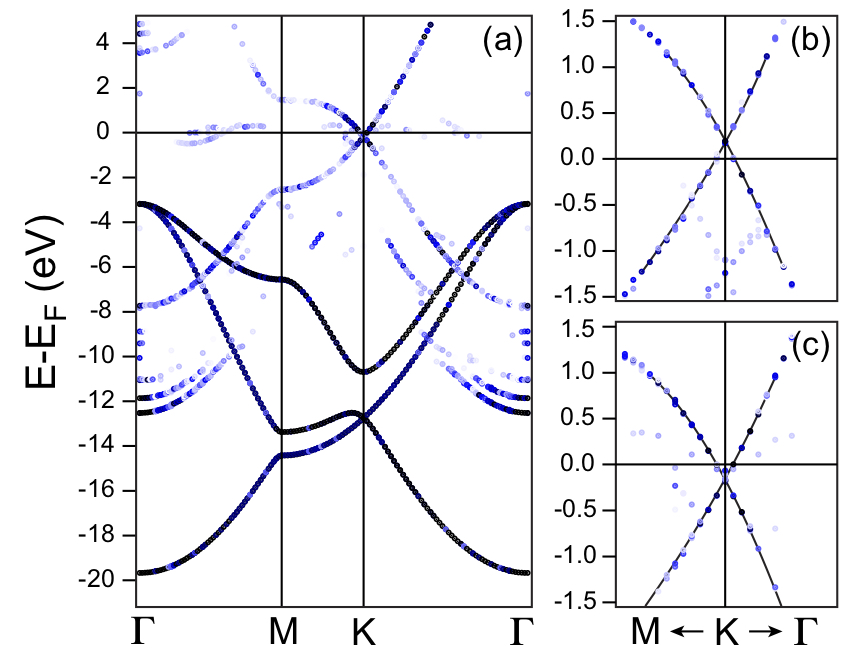}\\
\caption{(a) Calculated electron energy dispersion of the graphene-derived $\pi$ and $\sigma$ valence band states along the main directions of the hexagonal Brillouin zone for the gr/$n$-Ge(110) system with Sb dopants placed at the interface. The structures with the respective unit cell are shown in Fig.~\ref{grGe_STM_bias}(f) (graphene is $n$-doped). (b) Zoom of the energy dispersion of the graphene $\pi$ states in the vicinity of the $\mathrm{K}$-point for the clean gr/Ge(110) interface (graphene is $p$-doped). (c) Zoom of the energy dispersion of the graphene $\pi$ states in the vicinity of the $\mathrm{K}$-point from (a) (graphene is $n$-doped).}
\label{grGe_theory}
\end{figure}

For the clean gr/Ge(110) interface, without inclusion of any dopants, the equilibrium geometry with the simulated STM images is shown in Fig.~\ref{grGe_STM_bias}(e) (the armchair direction of the graphene lattice is parallel to the $<1\bar{1}0>$ direction of Ge(110)). According to the band structure calculations, graphene is $p$-doped in this system with the position of the Dirac point at $E_D-E_F=+195$\,meV (see panel (b) of Fig.~\ref{grGe_theory}), which is consistent with the previously published results for graphene grown on the Ge(001)/Si(001) epilayer~\cite{Dabrowski:2016im}. Generally, the band structure of graphene with the considered gr/Ge interface resembles the one of the free-standing graphene layer, but just shifted upwards due to doping and without any indication of orbital overlap of the valence band states of graphene and substrate. Our attempts to reproduce the correct doping of graphene observed in experiment via implantation of Sb atoms (according to the substrate specification) in the Ge slab does not improve the situation - graphene remains $p$-doped with only a small energy variation of the Dirac point position.

In order to reproduce the correct doping level of graphene on Ge(110), we based our next model for this interface on our STM results, which clearly indicate the presence of a large amount of dopant atoms at the gr/Ge(110) interface. [Here we would like to note that at the resulting gr/$n$-Ge(110) interface the positions of dopant Sb atoms is random and cannot be clearly identified. Therefore in our DFT calculations we limit ourselves to only one atom distribution (intercalation of Sb in gr/Ge), although the local variations of the dopant distributions, e.\,g. interstitial atoms in the Ge slab or/and in graphene, can locally vary the calculated doping level.] In the considered crystallographic models, the Sb atoms were regularly placed at the interface in different concentrations, i.\,e. the intercalation-like gr/Sb/Ge(110) system was considered. One of such equilibrium structures, when Sb atoms were initially placed directly under the carbon atoms (within the $(\sqrt{3}\times\sqrt{3})R30^\circ$ structure with respect to the graphene lattice) with a concentration of $27$ Sb atoms per $(9\times9)$ graphene supercell, and the respective simulated STM images are displayed in Fig.~\ref{grGe_STM_bias}(f). The resulting band structure of this system unfolded on the original $(1\times1)$ unit cell of graphene, i.\,e. presenting the weight of the $\pi$ and $\sigma$ states, is shown in Fig.~\ref{grGe_theory}(a) with the respective zoom around the $\mathrm{K}$-point in Fig.~\ref{grGe_theory}(c). From these results we conclude that graphene is $n$-doped in this system with a position of the Dirac point of $E_D-E_F=-170$\,meV, which is in a rather good agreement with the experimental ARPES and STS data. As was mentioned earlier, the exact modelling of such a disordered interface is very challenging as many irregular positions for Sb dopants have to be considered.

\section{Conclusion}

In conclusion, we performed detailed studies of the electronic structure of the graphene/$n$-Ge(110) interface from several $\mu\mathrm{m}$ down to nm-scale. This system was formed via atomic carbon deposition on hot Ge(110) and it is demonstrated that the substrate temperature during this procedure is the crucial factor defining the alignment of the formed graphene layer. The increase of the deposition temperature leads to the transition from two-domain to single-domain graphene growth on Ge(110). For both kinds of samples, two- and single-domain, graphene on Ge(110) is almost free-standing and no orbital overlap of the valence band states of graphene and Ge support is found. However, opposite to the previous experiments for graphene grown on the Ge(001)/Si(001) epilayers, in our ARPES and STM/STS experiments graphene is found $n$-doped with an $E_D$ position of $-210$\,meV and $-249$\,meV, respectively. This effect is explained by the formation of a disordered layer of Sb dopants at the gr/Ge(110) interface that was successfully confirmed by our DFT calculations. These dopants might appear at the Ge(110) surface during the preparation routine (sputtering/annealing) as well as during high temperature carbon deposition. Our low temperature spectroscopic experiments also demonstrate the Fermi velocity renormalization explained by the presence of the semiconducting substrate with the finite value of the dielectric constant, that leads to reduced screening of the electron-electron interaction at the gr/Ge interface. Our findings, on the one hand, demonstrate the perspectives on the application of the possible doping level tuning of graphene on semiconductors depending on the substrate treatment and kinds of dopants. On the other hand, such effects can drastically, and in some cases uncontrollably, change the properties of the graphene-semiconductor interfaces that might limit their applicability for future graphene-based semiconductor nano(micro)electronics and spintronics.

\section*{Methods}

\subsection*{Sample preparation}

The growth of graphene and the initial characterization by means of LEED and STM/STS were performed in an Omicron Cryogenic STM facility (base pressure $<1\times10^{-10}$\,mbar). Prior to every experiment a Ge(110) substrate (G-materials (Germany), Sb doped, resistivity $0.35\,\Omega\,\cdot\mathrm{cm}$) was cleaned via several cycles of Ar$^+$-sputtering ($1.5$\,keV, $p=1\times10^{-5}$\,mbar, $15$\,min) and annealing ($T=870^\circ$\,C, $10$\,min). Graphene was grown on the hot Ge(110) substrate at two different temperatures $T=850^\circ$\,C (Sample A) and $T=900^\circ$\,C (Sample B) from an atomic carbon source (Dr. Eberl MBE-Komponenten GmbH) with a filament current of $I=70$\,A and maximum pressure of $2\times10^{-9}$\,mbar during C-deposition. The cleanliness and quality of the prepared samples were controlled by LEED and STM. After preparation and STM/STS studies all samples were transferred under Ar-atmosphere to other experimental facilities for photoemission and microscopic studies. Prior to every one of the following experiments all samples were annealed under UHV conditions at $T=800^\circ$\,C for, at least, $30$\,min and the cleanliness and quality of the samples were controlled by core-level and valence band photoemission as well as LEED.      

\subsection*{Microscopy}

STM and STS measurements were performed at $\approx 10$\,K in an Omicron Cryogenic STM. Polycrystalline tungsten tips flash-annealed in UHV were used in all experiments. The sign of the bias voltage corresponds to the potential applied to the sample. Differential conductance ($dI/dV$) maps were recorded by means of standard \textit{lock-in} technique by applying a modulation voltage of $10$\,mV (rms) at a modulation frequency of $693.7$\,Hz to the tunneling voltage.
 
LEEM experiments were performed on a SPECS FE-LEEM P90 system at a base pressure of $2\times10^{-10}$\,mbar. The sample potential was $15$\,kV with respect to the objective lens.

\subsection*{Spectroscopy}

All photoelectron spectroscopy experiments were performed at the BESSY\,II storage ring (HZB Berlin). NEXAFS experiments were performed at the Russian-German beamline in the partial electron yield (PEY) mode (repulsive potential $U=-100$\,V) at the C $K$-edge with an energy resolution of $100$\,meV. Absorption experiments performed at different angles between sample surface and the direction of the incoming light allow to verify spatial orientation of the valence band states via the so-called \textit{search-light-like effect}.

The core-level XPS and valence-band ARPES experiments were performed at the UE\,56/2-PGM\,1 beamline. Sample A was measured in the photoemission station using a PHOIBOS\,100-2D-CCD hemispherical analyzer from SPECS GmbH. In these experiments, a 5-axis motorized manipulator was used, allowing for a precise alignment of the sample in $k$-space. The sample was pre-aligned (via polar and azimuth angles rotations) in such a way that the sample tilt scan was performed along the $<1\bar{1}0>$ direction of the Ge(110) substrate with the photoemission intensity on the channelplate images acquired along the $<00\bar{1}>$ direction of Ge(110). The final 3D data set of the photoemission intensity as a function of kinetic energy and two emission angles, $I(E_{kin},angle1,angle2)$, were converted into the $I(E_{B},k_x,k_y)$ sets and were then carefully analyzed.

Sample B was measured in the new experimental station equipped with a KREIOS\,150 electron analyzer from SPECS GmbH. The lens combination in this analyzer allows to collect electrons with the 2D-CCD detector in one $k_x$-direction for the full photoelectron emission hemisphere of $\pm90^\circ$. A hemispherical analyzer needs an entrance slit to select the $k_y$-vectors for energy dispersion. As a result, the second dimension in reciprocal space is scanned with a scanning-lens option in order to obtain a full 3D data set, $I(E_{kin},k_x,k_y)$. This analyzer also allows for real space photoelectron intensity imaging via scanning the lateral resolved 1D profile along the second dimension above the entrance slit. With that and 2D movable field apertures a well defined measurement position can be chosen. The sample is placed on the hexapod in this station allowing for its careful pre-alignment in $k$-space. The photon energies for all XPS and ARPES data sets are specified in the text and in the figure captions. All photoelectron spectroscopy measurements were performed at a sample temperature of $100$\,K. In all measurements, the correct position of the Fermi level was determined with a Mo-poly plate which was in contact with the gr/Ge(110) samples.

\subsection*{DFT calculations}

DFT calculations based on plane-wave basis sets of $500$\,eV cutoff energy were performed with the Vienna ab initio simulation package (VASP)~\cite{Kresse:1994,Kresse:1996}. The Perdew-Burke-Ernzerhof (PBE) exchange-correlation functional~\cite{Perdew:1996} was employed. The electron-ion interaction was described within the projector augmented wave (PAW) method~\cite{Blochl:1994} with C\,$(2s, 2p)$, Ge\,$(4s,4p)$, Sb\,$(4d, 5s, 5p)$ and H\,$(1s)$ states treated as valence states. The Brillouin-zone integration was performed on $\Gamma$-centred symmetry reduced Monkhorst-Pack meshes using a Gaussian smearing with $\sigma = 0.05$\,eV, except for the calculation of total energies. For those calculations, the tetrahedron method with Bl\"ochl corrections~\cite{Blochl:1994vg} was employed. A $6\times 6\times 1$ $k$-mesh was used. The DFT+$U$ scheme~\cite{Anisimov:1997gm,Dudarev:1998vn} was adopted for the treatment of Ge\,$2p$ orbitals, with the parameter $U_{eff}=U-J$ equal to $-3.5$\,eV~\cite{Persson:2006gy}. This approach reproduces all features of the \textit{fcc}-Ge band structure including an indirect band gap of $0.66$\,eV, the direct gap at $\Gamma$ of $0.94$\,eV and the spin-orbit splitting of $0.29$\,eV. For comparison: The calculations employing HSE06 (a hybrid functional)~\cite{Heyd:2003eg} give values of $0.63$\,eV, $0.73$\,eV, $0.29$\,eV, respectively~\cite{Peralta:2006bh,Stroppa:2011cb}. The corresponding experimental values are $0.66$\,eV, $0.89$\,eV and $0.29$\,eV~\cite{Grzybowski:2011im}. Dispersion interactions were considered adding a $1/r^6$ atom-atom term as parameterised by Grimme (``D2'' parameterisation)~\cite{Grimme:2006}. 

The gr/Ge(110) interface with the armchair edge of graphene parallel to Ge $\langle 1\bar{1}0\rangle$ was modelled by a slab consisting of five Ge layers, with a graphene layer adsorbed from top side of the slab and a vacuum gap of approximately $23$\,\AA. The used supercell has a ($9\times 9$) lateral periodicity with respect to the graphene layer and ($4\times 4$) lateral periodicity with respect to the unit cell of the Ge(110) surface. The dangling bonds appearing at the bottom side of the slab are substituted by H-atoms. In the case of gr/Sb/Ge(110), three concentrations were investigated: $1$\,ML ($27$ Sb atoms per ($9\times 9$) graphene-related supercell), $0.6$\,ML, and $0.15$\,ML. In each situation, Sb atoms were evenly distributed at the interface between graphene and Ge(110). In order to model the gr/Ge$_x$Sb$_y$ system, four Ge atoms of the interface layer or of the middle Ge-layer, respectively, were substituted by the Sb-atoms. During the structural optimisation procedure, the four bottom layers (three Ge layers and a H-layer) were fixed at their bulk positions and the Ge-H distance of $1.54$\,\AA, whereas the positions ($x,\,y,\,z$-coordinates) of all other ions were fully relaxed until forces became smaller than $0.02\,\textrm{eV\,\AA}^{-1}$. 

The band structures calculated for the studied systems were unfolded to the graphene ($1\times 1$) primitive unit cells according to the procedure described in Refs.~\citenum{Medeiros:2014ka,Medeiros:2015ks} with the code BandUP. The STM images are calculated using the Tersoff-Hamann formalism~\cite{Tersoff:1985}.

\section*{Acknowledgement}

The authors thank the German Research Foundation (DFG) for financial support within the Priority Programme 1459 ``Graphene''. We acknowledge the North-German Supercomputing Alliance (HLRN) for providing computer time. This research used resources of the Advanced Light Source, which is a DOE Office of Science User Facility under contract no. DE-AC02-05CH11231.



\providecommand{\latin}[1]{#1}
\providecommand*\mcitethebibliography{\thebibliography}
\csname @ifundefined\endcsname{endmcitethebibliography}
  {\let\endmcitethebibliography\endthebibliography}{}



\clearpage

\clearpage

\clearpage

\clearpage

\clearpage

\clearpage

\clearpage

\clearpage
\noindent
Supplementary material for manuscript:\\

\noindent
\textbf{The graphene/$n$-Ge(110) interface: structure, doping, and electronic properties}\\
\newline
Julia Tesch,$^1$ Fabian Paschke,$^1$ Mikhail Fonin$^1$, Marko Wietstruk,$^2$ Stefan B\"ottcher,$^2$ Roland J. Koch,$^3$ Aaron Bostwick,$^3$ Chris Jozwiak,$^3$ Eli Rotenberg,$^3$ Anna Makarova,$^4$ Beate Paulus,$^5$ Elena Voloshina,$^{6,*}$ and Yuriy Dedkov$^{6,**}$\\
\newline
$^1$~{Fachbereich Physik, Universit\"at Konstanz, 78457 Konstanz, Germany}\\
$^2$~{SPECS Surface Nano Analysis GmbH, Voltastra\ss e 5, 13355 Berlin, Germany}\\
$^3$~{Advanced Light Source, E. O. Lawrence Berkeley National Laboratory, Berkeley, CA, 94720, USA}\\
$^4$~{Institut f\"ur Festk\"orperphysik, Technische Universit\"at Dresden, 01062 Dresden, Germany}\\
$^5$~{Institut f\"ur Chemie und Biochemie, Freie Universit\"at Berlin, Takustra\ss e 3, 14195 Berlin, Germany}\\
$^6$~{Department of Physics and International Centre for Quantum and Molecular Structures, Shanghai University, Shangda Road 99, 200444 Shanghai, China}\\
$^*$~{Corresponding author. E-mail: voloshina@shu.edu.cn}\\
$^{**}$~{Corresponding author. E-mail: dedkov@shu.edu.cn}\\

\clearpage

\noindent\textbf{Movie\,S1.} (available on request) The movie shows the binding energy scan through the valence band of the graphene/$n$-Ge(111) system for Sample A (two-graphene-domains sample). Left panel: Constant energy cuts at different binding energies; Brillouin zones with different $\mathrm{K}$-points, $\mathrm{K}_1$ and $\mathrm{K}_2$, corresponding to two different graphene domains are marked by two dashed-line hexagons. Right panel: a series of the ARPES spectra along the $\Gamma-\mathrm{K}_1$ direction (marked in the left panel) of the hexagonal graphene-derived Brillouin zone corresponding to one of the graphene domains; $\pi_1$ and $\pi_2$ states are marked in the plot. Data were collected with the PHOIBOS\,100/2D-CCD analyzer and photon energy of $h\nu=100$\,eV.\\

\noindent\textbf{Movie\,S2.} (available on request) The movie shows the binding energy scan through the valence band of the graphene/$n$-Ge(111) system for Sample B (single-graphene-domain sample). Left panel: Constant energy cuts at different binding energies; Brillouin zone of graphene is marked by dashed-line hexagon. Right panel: a series of the ARPES spectra along the $\mathrm{K}-\Gamma-\mathrm{K}$ direction (marked in the left panel) of the hexagonal graphene-derived Brillouin zone. Data were collected with the KREIOS\,150/2D-CCD analyzer and photon energy of $h\nu=100$\,eV.\\

\clearpage

\noindent\textbf{Table\,S3.} A set of parameters obtained for different gr-Ge systems discussed in the text: graphene layer corrugation (in \AA), mean distance between a graphene layer and top Ge layer (in \AA), mean distance between layer of Sb dopants and top Ge layer for the intercalation-like systems (in \AA), interaction energy (in meV per C-atom), position of the Dirac point with respect to the Fermi level (in meV). Considered systems: 1 - clean gr/Ge(110) interface (Fig.\,S4); 2 - gr/Ge(110) where $4$ Sb atoms replace random Ge atoms in the top layer (Fig.\,S5); 3 - same as 2 but $4$ Sb atoms replace Ge atoms in the 3rd Ge layer (Fig.\,S6); 4 - gr/Sb/Ge(110) intercalation-like system with concentration of $4$ Sb atoms per $(9\times9)$ graphene supercell (Fig.\,S7); 5 - gr/Sb/Ge(110) intercalation-like system with concentration of $16$ Sb atoms per $(9\times9)$ graphene supercell (Fig.\,S8); 6,7 - gr/Sb/Ge(110) intercalation-like system with concentration of $27$ Sb atoms per $(9\times9)$ graphene supercell, before geometry optimization Sb atoms were placed either directly under C-atoms (6, Fig.\,S9) or in the center of the C-ring (7, Fig.\,S10).\\

\begin{tabular}{|l|c|c|c|c|c|}
\hline
       		 &gr-corr.  &gr-Ge dist.  &Sb-Ge dist. & $E_\textrm{int}$  &$E_D-E_F$	\\
System 		 &(\AA)     &(\AA)        &(\AA)       &(meV/C-atom) &(meV)	\\[5pt]
\hline
1: gr/Ge(110)      	&$0.09$	&$3.54$ &  	&$-42$  &$+195$  \\[5pt]
\hline
2: gr/Ge$_x$Sb$_y$ 	&$0.14$	&$3.56$ &  	&$-42$  &$+100$  \\
(4 Sb in layer 1)	&  	&  	&  	&  	&   \\[5pt]
\hline
3: gr/Ge$_x$Sb$_y$  	&$0.09$	&$3.52$ &  	&$-42$  &$+165$  \\
(4 Sb in layer 3)	&  	&  	&  	&	&  \\[5pt]
\hline
4: gr/Sb/Ge	 	&$0.49$	&$4.03$ &$1.27$	&$-32$  &$\,\,\,+45$  \\
(4\,Sb/u.c.) 	 	&  	&  	&  	&  	&  \\[5pt]
\hline
5: gr/Sb/Ge  	 	&$0.42$	&$4.91$ &$1.47$ &$-33$  &$\,\,\,-95$  \\
(16\,Sb/u.c.) 	 	&  	&  	&  	&  	&  \\[5pt]
\hline
6: gr/Sb/Ge  	 	&$0.43$ &$5.50$ &$2.03$ &$-41$  &$-170$  \\
(27\,Sb/u.c.) 	 	&  	&  	&  	&  	&  \\[5pt]
\hline
7: gr/Sb/Ge  	 	&$0.43$ &$5.46$ &$1.96$ &$-41$  &$-125$  \\
(27\,Sb/u.c.) 	 	&  	&  	&  	&  	&  \\[5pt]
\hline
\end{tabular}

\clearpage

\begin{figure}
\includegraphics[width=\textwidth]{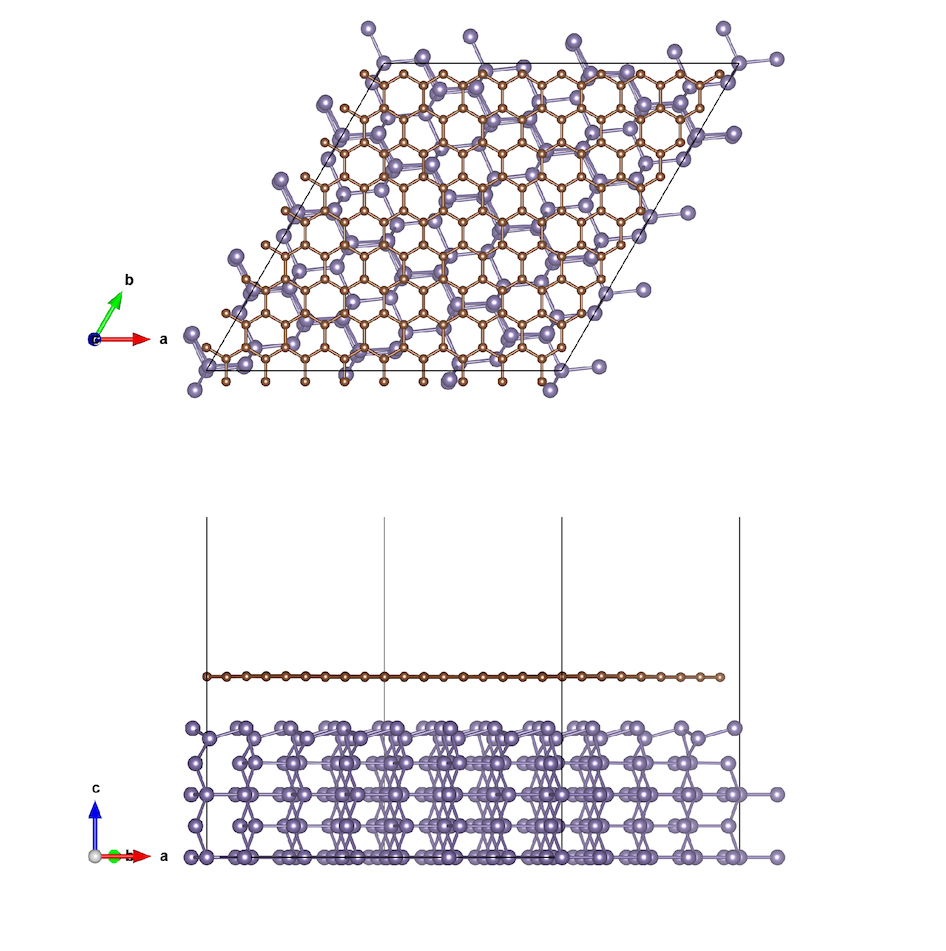}
\end{figure}
\noindent\textbf{Figure\,S4.} Top and side views of the clean gr/Ge(110) interface after geometry optimization (System 1, Table S3).

\clearpage

\begin{figure}
\includegraphics[width=\textwidth]{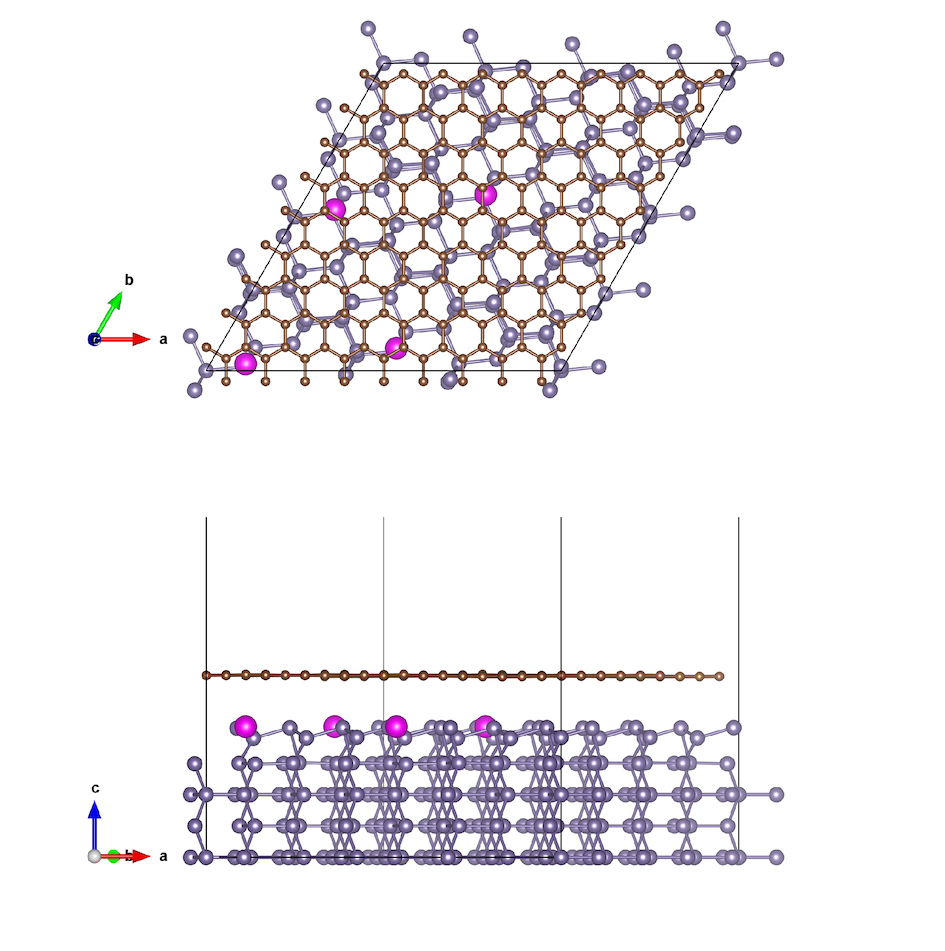}
\end{figure}
\noindent\textbf{Figure\,S5.} Top and side views of the gr/Ge(110) interface after geometry optimization where 4 Sb atoms replace random Ge atoms in the top layer (System 2, Table S3).

\clearpage

\begin{figure}
\includegraphics[width=\textwidth]{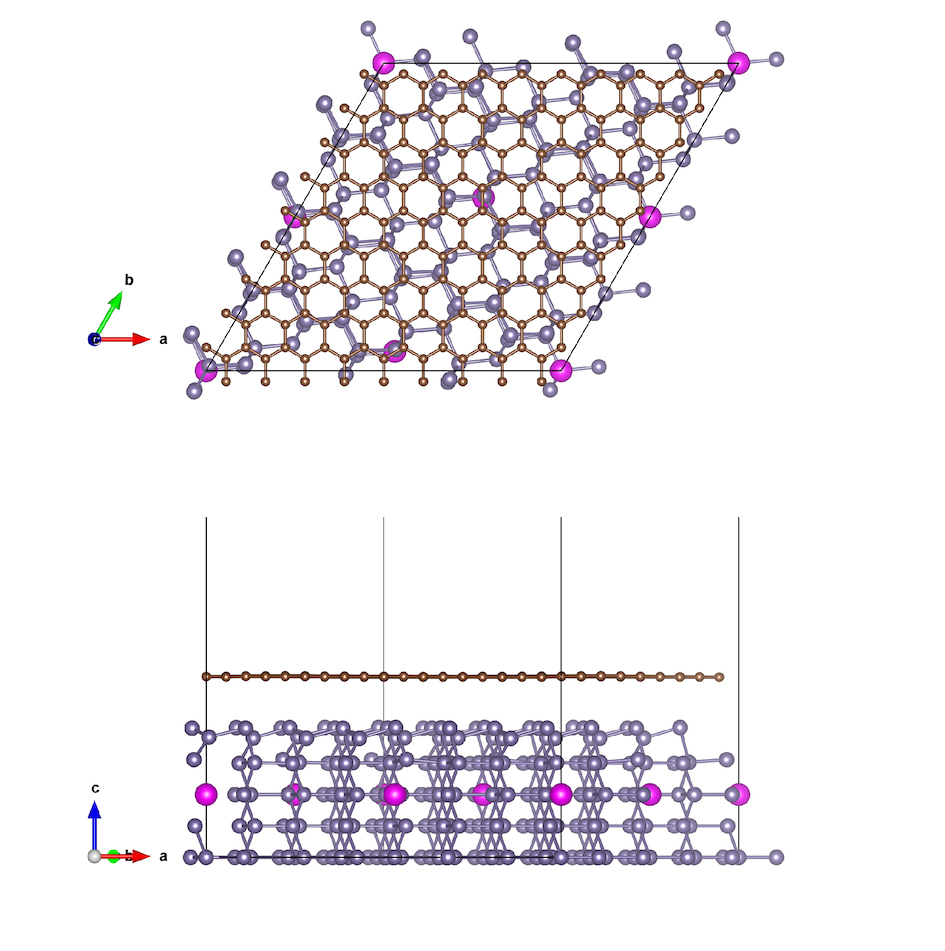}
\end{figure}
\noindent\textbf{Figure\,S6.} Top and side views of the gr/Ge(110) interface after geometry optimization where 4 Sb atoms replace random Ge atoms in 3rd layer (System 3, Table S3).

\clearpage

\begin{figure}
\includegraphics[width=\textwidth]{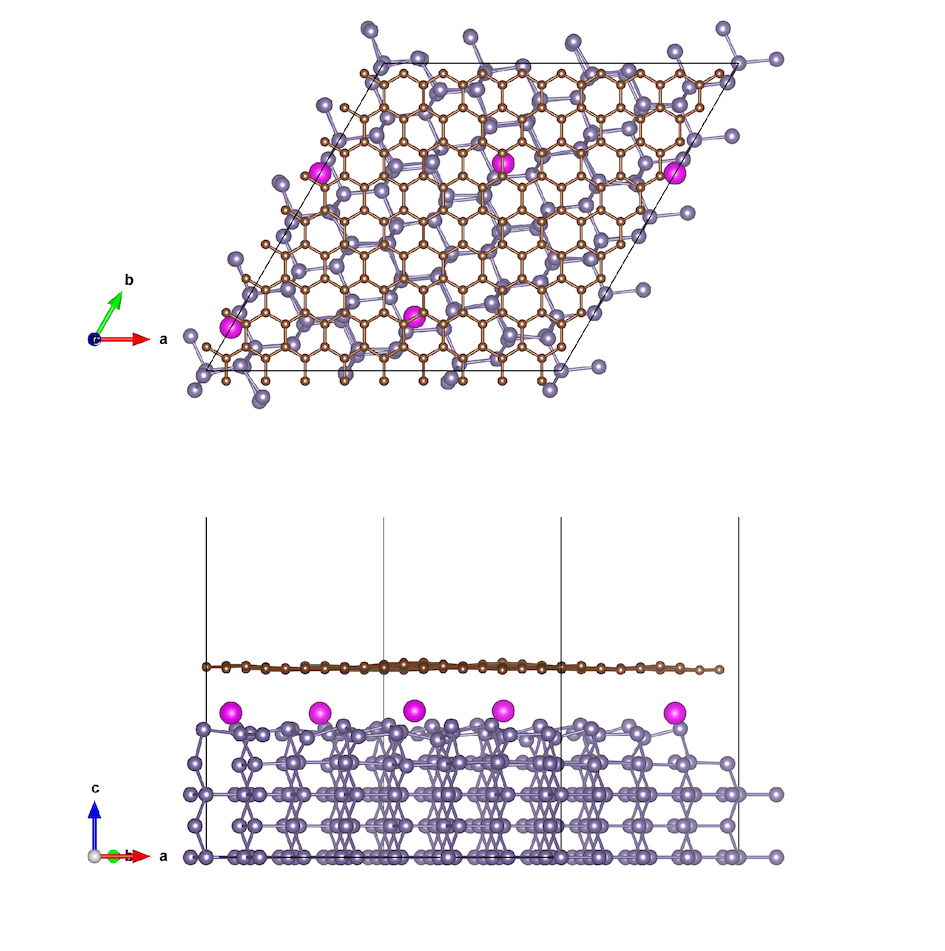}
\end{figure}
\noindent\textbf{Figure\,S7.} Top and side views of the gr/Sb/Ge(110) intercalation-like system after geometry optimization with concentration of $4$ Sb atoms per $(9\times9)$ graphene supercell (System 4, Table S3).

\clearpage

\begin{figure}
\includegraphics[width=\textwidth]{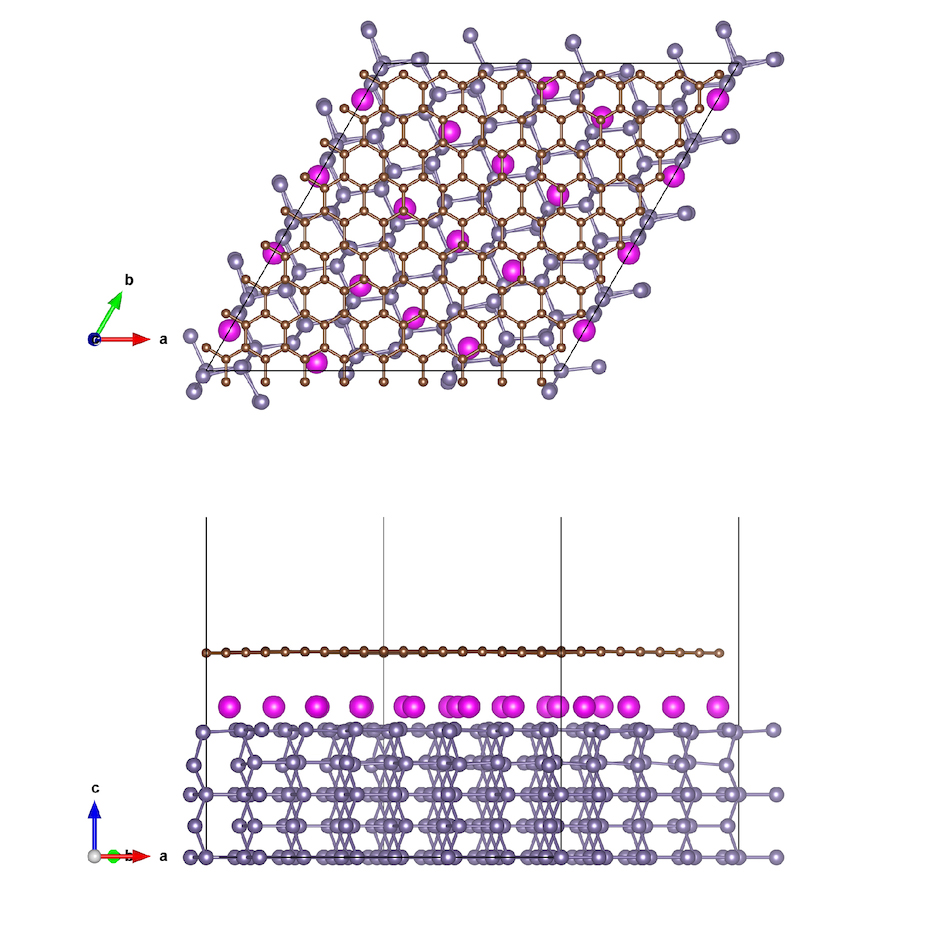}
\end{figure}
\noindent\textbf{Figure\,S8.} Top and side views of the gr/Sb/Ge(110) intercalation-like system after geometry optimization with concentration of $16$ Sb atoms per $(9\times9)$ graphene supercell (System 5, Table S3).

\clearpage

\begin{figure}
\includegraphics[width=\textwidth]{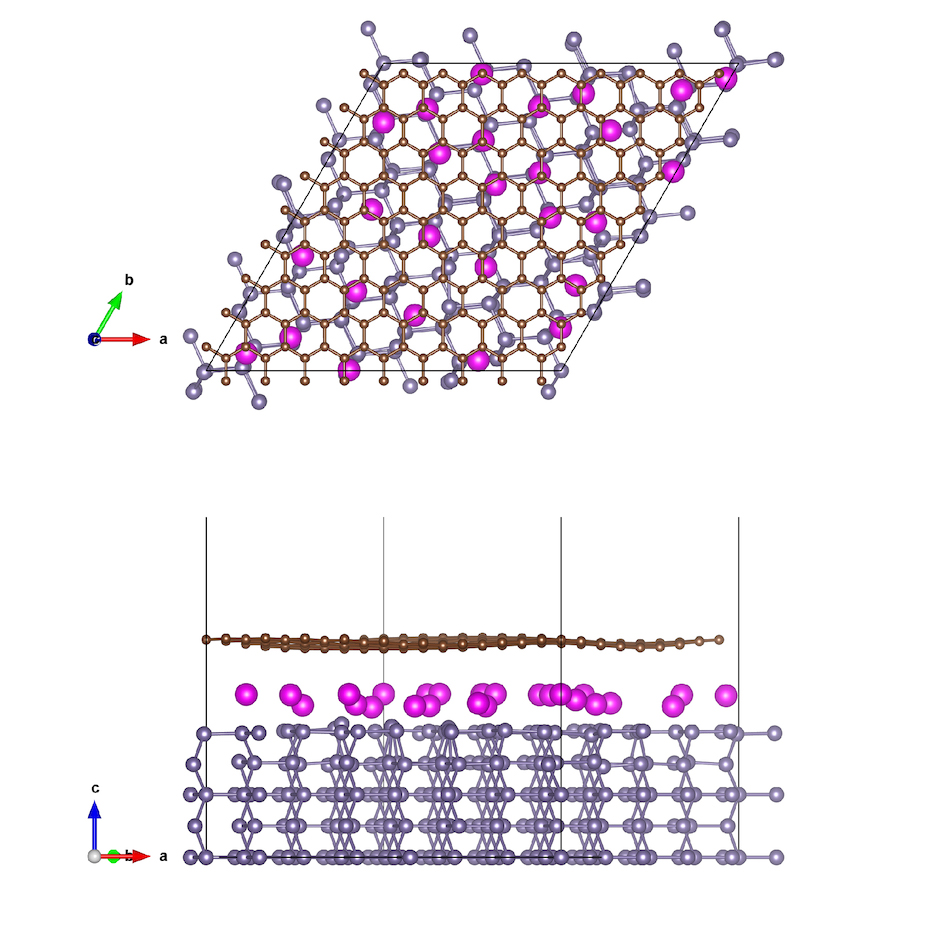}
\end{figure}
\noindent\textbf{Figure\,S9.} Top and side views of the gr/Sb/Ge(110) intercalation-like system after geometry optimization with concentration of $27$ Sb atoms per $(9\times9)$ graphene supercell (System 6, Table S3). Before geometry optimization Sb atoms were placed directly under C-atoms.

\clearpage

\begin{figure}
\includegraphics[width=\textwidth]{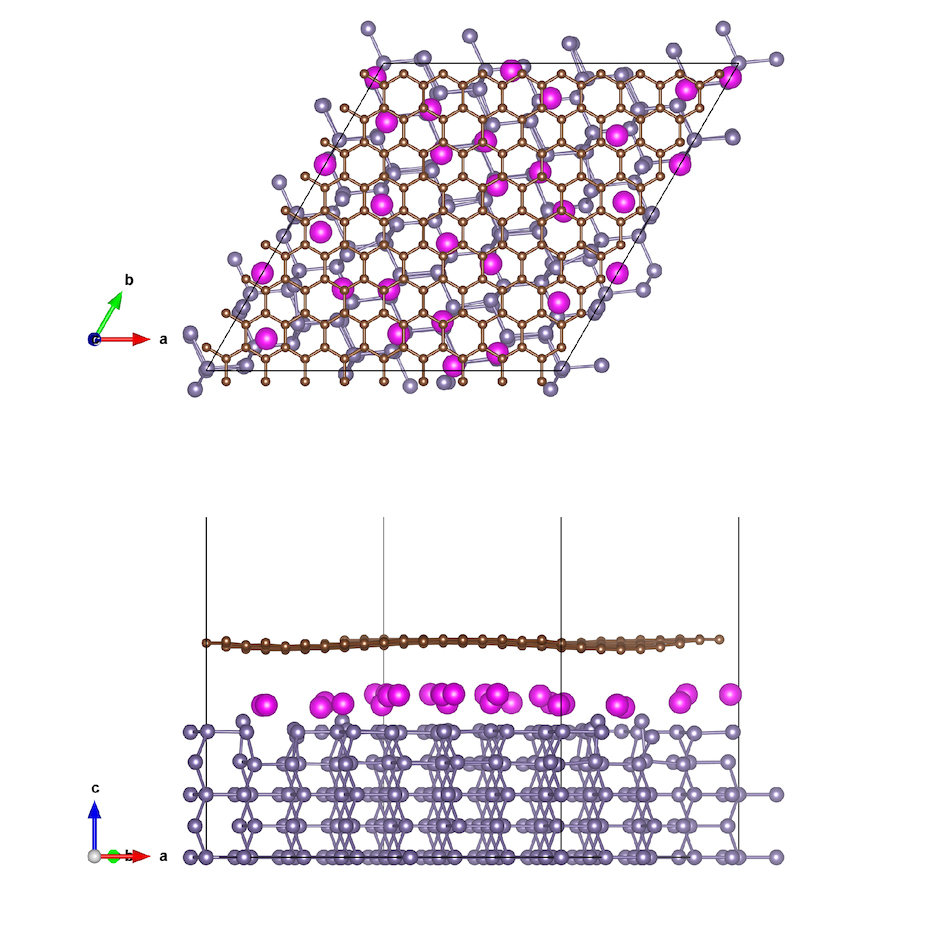}
\end{figure}
\noindent\textbf{Figure\,S10.} Top and side views of the gr/Sb/Ge(110) intercalation-like system after geometry optimization with concentration of $27$ Sb atoms per $(9\times9)$ graphene supercell (System 7, Table S3). Before geometry optimization Sb atoms were placed in the center of the C-rings.

\end{document}